# Conservation of orbital angular momentum in air core optical fibers

P. Gregg,[1] P. Kristensen,[2] and S. Ramachandran[1,*]

[1]*Boston University, 8 St. Mary's St, Boston, MA, USA*
[2] *OFS-Fitel, Priorparken 680, Brøndby 2605, Denmark*
*\*Corresponding author*: sidr@bu.edu

(3 December 2014)

Light's orbital angular momentum (OAM) is a conserved quantity in cylindrically symmetric media; however, it is easily destroyed by free-space turbulence or fiber bends, because anisotropic perturbations impart angular momentum. We observe the conservation of OAM even in the presence of strong bend perturbations, with fibers featuring air cores that appropriately sculpt the modal density of states. In analogy to the classical reasoning for the enhanced stability of spinning tops with increasing angular velocity, these states' lifetimes increase with OAM magnitude. Consequently, contrary to conventional wisdom that ground states of systems are the most stable, OAM longevity in air-core fiber increases with mode-order. Aided by conservation of this fundamental quantity, we demonstrate fiber propagation of 12 distinct higher-order OAM modes, of which 8 remain low-loss and >98% pure from near-degenerate coupling after km-length propagation. The first realization of long-lived higher-order OAM states, thus far posited to exist primarily in vacuum, is a necessary condition for achieving the promise of higher-dimensional OAM-based classical and quantum communications over practical distances.

Quantum numbers are usually assigned to conserved quantities; hence it appears natural that paraxial light travelling in isotropic, cylindrically symmetric media such as free space or optical fibers be characterized by its angular momentum [1]:

$$\mathcal{J} = \mathcal{L} + \mathcal{S} \qquad (1)$$

$\mathcal{L}$ represents light's orbital angular momentum (OAM) [2], and $\mathcal{S}$ represents its spin angular momentum, SAM, commonly known as left or right handed circular polarization, $\hat{\sigma}^{\pm}$, such that $\mathcal{S} = \pm 1$ in units of $\hbar$ for left and right handed circular polarization, respectively. $\mathcal{L}$ forms a countably infinite dimensional basis, spawning widespread interest in OAM beams [3,4,5]. In particular, this enables a large alphabet of states for hyper-entangled quantum communications or high-capacity classical links. The information capacity of a classical or quantum communications link increases with the number of distinct, excitable and readable orthogonal information channels. Degrees of freedom that conserve their eigenvalues are required, because perturbations which cause eigenstate rotation, conventionally called mode coupling, are debilitating. In classical communications, computational algorithms can partially recover information for some limited perturbations, albeit with energy-intensive signal processing [6]. For low-light level applications such as quantum communications or interplanetary links, the information is lost. With the use of wavelength and polarization as photonic degrees of freedom virtually exhausted, the recent past has seen an explosion of interest in a new degree of freedom – orthogonal spatial modes that are stable during propagation, of which OAM is one interesting choice [7,8].

In practice, this choice of quantum numbers is questionable. Although large ensembles of OAM modes can be generated [7,8, 9,10,11], they are easily destroyed by anisotropic perturbations such as atmospheric turbulence [12] in free space, or bends in fibers [13], limiting OAM transmission experiments to primarily laboratory length scales (meters) [8,14]. Practical communications distances, over fiber or free-space, have been achieved only for special case of the lowest-order (|L|=1) OAM states [15,16]. OAM transmission is hampered by near-degeneracies of the desired OAM state with a multitude of other modes [17,18] possessing different |$L$| or radial quantum numbers. These near-degeneracies in linear momentum, or equivalently longitudinal wavevector, $k_z$, (in waveguides, also represented by effective index, neff, given by $k_z = 2\pi \cdot n_{eff}/\lambda$, where $\lambda$ is the free space wavelength, and z signifies propagation coordinate), phase match orthogonal modes and couple them in the presence of perturbations. Since any multimodal system would, by definition, have a high density of states, this is a fundamental problem, and exploiting the infinite-dimensional basis afforded by OAM beams requires a medium in which this modal degeneracy is addressed.

Here, we report the design of a general class of optical fibers, featuring an air core, that enables conservation of OAM (**Fig. 1**). The air core acts as a repulsive barrier, forcing the mode field to encounter the large index step between ring and cladding (**Fig. 1b**). This lifts polarization near-degeneracies [19] of OAM states with the same |L|, though the states with OAM and SAM aligned (of the same handedness) remain degenerate with each other, but separate from those with OAM and SAM anti-aligned (**Fig. 1c**; more information in Supplementary §2). This $n_{eff}$ splitting generally increases with |L| (**Fig. 1d**). A key feature of the air core fiber [20] is the existence of modes with large |L| but the prevention of modes with a large radial quantum number whose neff may be close to the desired OAM states, by appropriate sculpting of mode volume to vastly decrease the density of states, (see Supplementary §4). The effect is similar to the restriction of the mode structure in microtoroid resonators, in which devices preferentially support equatorial modes [21].

Using the experimental apparatus [22] in **Fig. 2a**, we excite and propagate 12 OAM states over 10m of our air core fiber at 1530nm. Fiber output fields are imaged onto a camera through a circular polarization beam splitter, separating $\hat{\sigma}^+$ and $\hat{\sigma}^-$ into the right and left bins, respectively. Excitation of, for example, |L|=7 modes yields clean ring-like intensities which remain in the circular polarization selected by the quarter wave plate before the fiber. As the radial envelopes of the modes are nearly identical, we interfere with an expanded Gaussian beam (**Fig. 2a** orange path) to reveal their phase structure. For each interference pattern, the number of spiral arms (parastichies) indicates the mode's |$L$|, while the handedness (right or left) indicates the sign of $L$. Combined with sorting by circular polarization, we unambiguously identify OAM states. Clean spiral images (see Supplementary §5) in **Fig. 2b** indicate negligible coupling amongst, and hence clean transmission of, all 12 OAM states.

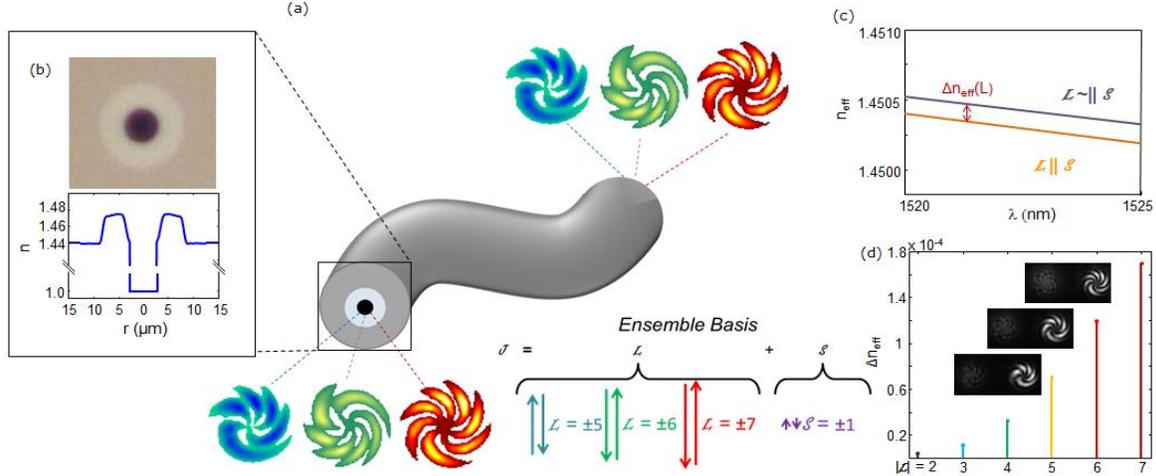

Fig. 1(a) Free space OAM states are coupled into air core fiber and conserved despite bends and random fiber-shape deformations. Each state possesses total angular momentum, $\mathcal{J}$, which comprises orbital, $\mathcal{L}$, and spin, $\mathcal{S}$, parts, which may be positive or negative (right or left-handed). Thus, there are 4 OAM states for every |L|, each of which can carry information. (b) Microscope image (top) and measured refractive index profile (bottom) for air core fiber supporting 12 OAM states. (c) Example of $n_{eff}$ splitting amongst OAM states. In conventional fibers, states of the same $|\mathcal{L}|$ are near-degenerate and freely couple. Via the air core design this near-degeneracy is broken such that states with spin and OAM aligned separate from states with spin and OAM anti-aligned. (d) Effective index splitting in a typical air core fiber: ~$10^{-4}$ is considered sufficient for OAM-propagation [13] achieved in this design for $|\mathcal{L}|$=5, 6, and 7.

The above result is counterintuitive – while the air-core fiber design lifts degeneracies amongst a host of OAM states, the modes still appear in degenerate pairs of spin-orbit aligned or anti-aligned states. The coefficient of power coupling between modes $j$ and $k$ is given by [17]:

$$\langle P_{j,k} \rangle = k^2 \Phi(k_j - k_k) \left( \iint r dr d\varphi \Delta n^2(r,\varphi) \psi_j^* \psi_k \right)^2 \quad (2)$$

$\psi_j$ and $k_{z,j}$ are the normalized electric field and longitudinal wavevector of the $j^{th}$ mode, respectively, $\Delta n^2$ is the index perturbation, and $\Phi(k_j - k_k)$ incorporates the perturbation's longitudinal behavior and is typically maximized for $k_{z,j} = k_{z,k}$ (see Supplementary §3). Thus, pairs of degenerate modes should be susceptible to coupling within their two-mode subspace via anisotropic perturbations such as the bends that existed on the 10-m long fiber. In fact, for lower order, $|\mathcal{L}| = 1$, OAM states, such coupling is possible [23] and controllably exploited [24] using a series of fiber loops, in analogy to a conventional polarization controller (polcon) in single-mode fiber (SMF). The polcon may be understood as transfer of OAM from the bend-induced perturbation to the field itself [25]. Any z-independent anisotropic perturbation may be expanded as:

$$\Delta n^2(r,\varphi) = \sum_{p=-\infty}^{\infty} a_p(r) e^{ip\varphi} \quad (3)$$

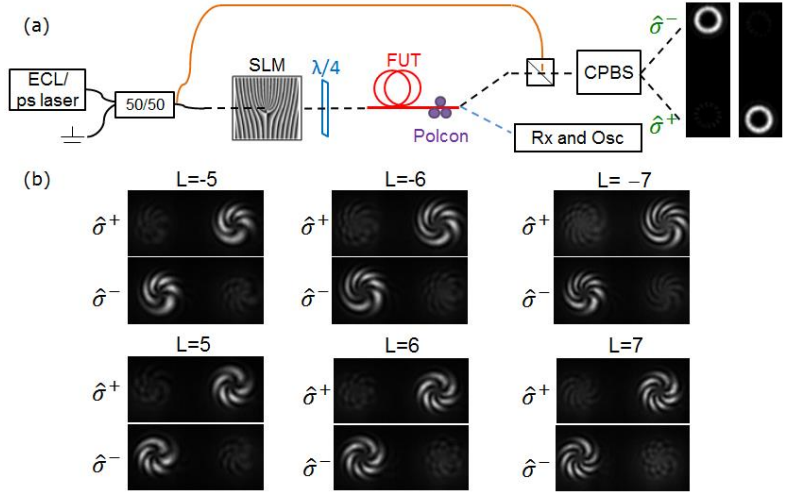

Fig. 2(a) Setup: light from an external cavity laser (ECL; 1530nm) or picosecond pulsed laser (1550nm) is converted to free space OAM modes via an SLM followed by a quarter wave plate, and launched into the air core fiber. The fiber's output is imaged through a circular polarization beam splitter (CPBS), separating $\hat{\sigma}^-$ from $\hat{\sigma}^+$. For interference measurements, source light tapped from a 50/50 splitter at the input (orange path) acts as a reference. For stability measurements, the air-core fiber passes through a polcon. For time of flight measurements (blue path), a fast detector and oscilloscope (Rx and Osc) are used. Dashed lines indicate free space, solid lines indicate fiber. Fiber output images are for $|L|=7$. (b) OAM states after 10m of the air core fiber, interfered with an expanded Gaussian reference. Text around images indicates launch conditions. 12 states for $|L|=5, 6, 7$ in all SAM/OAM combinations are shown. See Supplementary §6 for additional experimental details.

where $a_p(r)$ is the Fourier series coefficient of the perturbation corresponding to angular momentum $p \cdot \hbar$. Coupling from a mode with OAM $L_1$ to one with OAM $L_2$ depends on the inner product between the initial field, $\psi_1 = F_1(r)e^{iL_1\varphi}$, the perturbation, and the second field, $\psi_2 = F_2(r)e^{iL_2\varphi}$. Evaluating the angular part of this overlap integral:

$$\langle\psi_1|\Delta n^2(r,\varphi)|\psi_2\rangle = \sum_{p=-\infty}^{\infty}\langle F_1(r)|a_p(r)|F_2(r)\rangle\langle e^{iL_1\varphi}|e^{ip\varphi}|e^{iL_2\varphi}\rangle \quad (4)$$

yields the selection rule:

$$p - (L_1 - L_2) = 0 \quad (5)$$

Bends and shape deformations additionally induce birefringence, which couples spins, as does a conventional polcon for SMF [26]. Allowing for spin coupling, transitioning between degenerate higher order states requires a perturbation element of order $p = 2|L|$, which becomes increasingly negligible for large $|L|$ (**Fig. 3a**). To experimentally interrogate this curious effect, we build a polcon [27], but with the air core fiber (purple circles in **Fig. 2a**), with bend radius ~ 3cm. We define a degradation factor, $\alpha$, as the ratio of the maximum power in $\hat{\sigma}^+$ to that in $\hat{\sigma}^-$ when $\hat{\sigma}^+$ is launched, or vice versa. For high $|L|$ states, such as $L = -7\ \hat{\sigma}^+$ (**Fig. 3b**) degradation factors are typically < 10%. As expected, for the $L = 0$ mode in SMF (**Fig. 3c**), $\alpha \cong 1$, indicating complete coupling between the two degenerate SAM states. Due to the rapid decrease of $|a_p|$ as $L$ increases, the observed degradation factor decreases, with ratios as low as –12 dB (~6%) for higher |$\mathcal{L}$| states relative to SMF (**Fig. 3d**). Thus, we find that, for high |$\mathcal{L}$| states in air-core fibers, OAM is truly a conserved quantity even in the presence of anisotropic perturbations, since transitions amongst degenerate states are forbidden, based on conservation of OAM (**Fig. 3e**). This behavior parallels forbidden transitions between electron spin states with an externally applied electric field. Here, anisotropic bends assume the role of electric field perturbations, leaving the initial state unchanged.

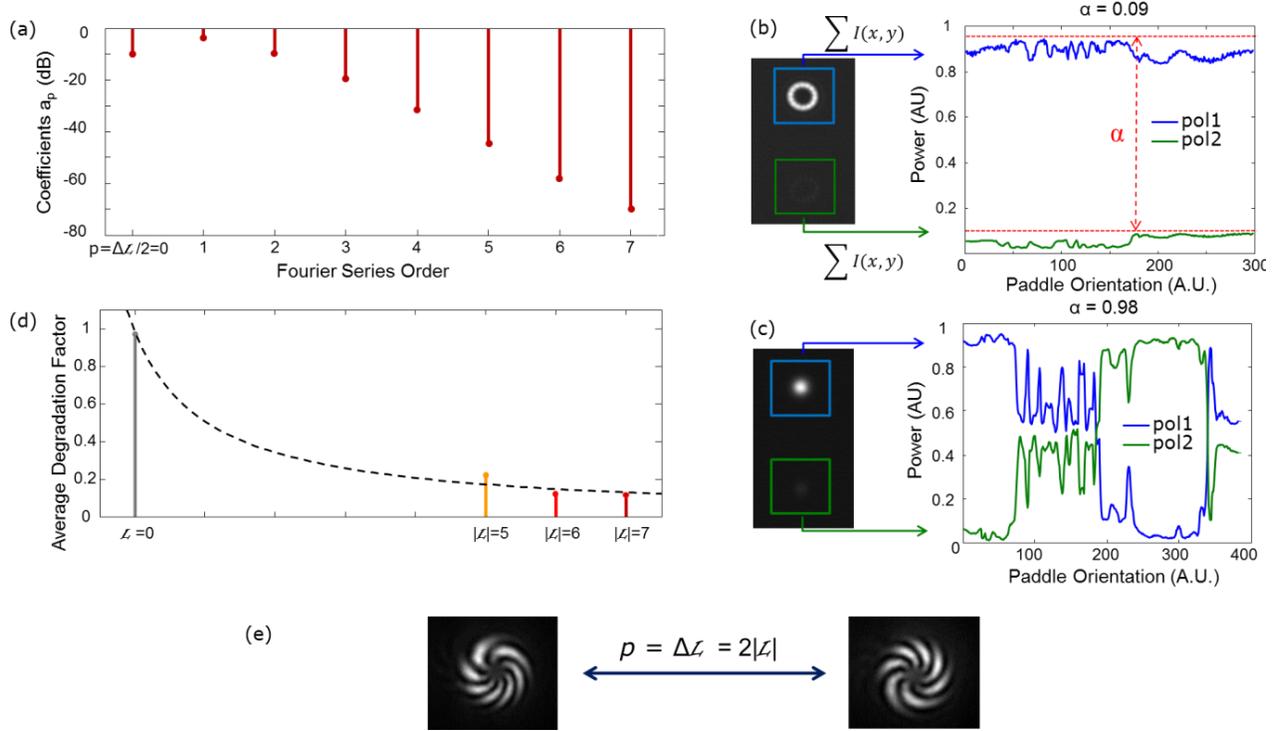

Fig. 3(a) Theoretical prediction, via Fourier series expansion, of the OAM content of a bend perturbation for an air core fiber. Coefficients rapidly decrease with increasing OAM content, $p$. (b) Illustration of power binning measurement for $\mathcal{L} = 7, \hat{\sigma}^+$. As the polcon paddles (**Fig. 2a**) are tuned, negligible coupling from $\hat{\sigma}^+$ to $\hat{\sigma}^-$ is observed, indicating degenerate-state stability. Legend "pol 1" indicates launched polarization, "pol 2" indicates parasitic polarization. (c) Polcon measurement for SMF indicating complete degenerate state mode coupling. (d) Experimentally measured average values of degradation factor, $\alpha$, for each |$\mathcal{L}$|, plotted against a shifted $1/|\mathcal{L}|$ trend line (dashed line). Degradation drops with increasing OAM. This concept was tested experimentally on states for which spin-orbit aligned to spin-orbit anti-aligned coupling is suppressed. (e) Schematic indicating the perturbation OAM content necessary to couple degenerate fiber states with opposite values of $\mathcal{L}$.

Over long enough interaction lengths, light may encounter other perturbation symmetries due to twists and imperfections in the draw process. We experimentally study long-length propagation by transmitting OAM states with a picosecond pulsed (70GHz bandwidth) laser at 1550nm (**Fig. 2a**) and measuring time-of-flight traces. At the output of fiber length $z$, modes j and k are temporally separated by $\Delta t^{j,k} = \Delta n_g^{j,k} z/c$. As all of the OAM modes in this fiber have similar group-velocity dispersions, relative mode purity, $\alpha$, conventionally referred to as multipath interference (MPI) [28], is given by:

$$\alpha = 10 \log \left( \frac{P_{peak\ 1} - \bar{P}_{noise}}{P_{peak\ 2} - \bar{P}_{noise}} \right) \tag{6}$$

where $\bar{P}_{noise}$ is the average noise power and Ppeak k is the peak power of the kth mode. Combined time-of-flight measurements for modes in the |L| =5, 6 families are shown in Fig. 4a and Fig. S4a, with close-ups of individual traces shown in Fig. 4(b-e) and Fig. S4(b-e). We find that MPIs of −18dB or greater (>98% purity) can be achieved for any |L| =5, 6 mode relative to the background, the |L| = 7 modes being too lossy for 1km transmission at 1550nm, due to fabrication imperfections. We obtain similar results over the wavelength range of 1530 to 1565 nm, thus confirming that the OAM states are wavelength-agnostic. Loss for the |$L$| =5 and 6 mode groups, measured via fiber-cutback, is 1.9 and 2.2 dB/km, respectively. Note that this only measures cross-coupling between spin-orbit aligned and anti-aligned states, as the degenerate states have identical group delays. When OAM states are propagated over km lengths, we observed $\hat{\sigma}^+$ to $\hat{\sigma}^-$ transitions at the fiber output. This potentially indicates twist perturbations, known to affect OAM stability [29]. Extending the quantum-mechanical analogy, twists would assume the role of magnetic perturbations, which couple electronic spin states. However, this coupling constitutes a unitary transformation within the 2-mode subspace and may be disentangled with devices such as q-plates [30], thus still yielding a medium in which all 8 of the states may be used for information encoding.

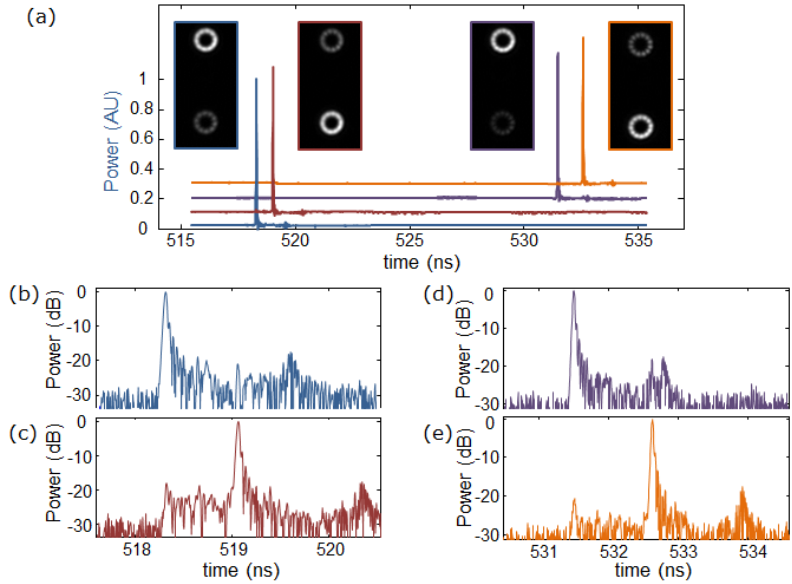

Fig. 4(a) Time of flight measurements, using setup of **Fig. 3a**, for four different OAM states. Traces vertically offset for visual clarity, in order of increasing group delay: $L$ = +5 $\hat{\sigma}^-$, $L$ = +5 $\hat{\sigma}^+$, $L$ = +6 $\hat{\sigma}^-$, and $L$ = +6 $\hat{\sigma}^+$. Inset: fiber output image after 1km propagation. (b) Close-up of time domain trace for spin-orbit anti-aligned $L$ = +5 $\hat{\sigma}^-$ mode (peak around 519.5ns is spurious from the detector's electrical impulse response). (c) Close-up of time domain trace for spin-orbit aligned $L$ = +5 $\hat{\sigma}^+$. The time difference between the two $L$ = +5 peaks, 0.75ns, corresponds well to the theoretical value of 0.7ns. (d) and (e) show close-ups of $L$ = +6 $\hat{\sigma}^-$ and $L$ = +6 $\hat{\sigma}^+$ traces, indicating even better parasitic mode suppression. The peaks from (d) and (e) would overlap in conventional step-index fibers due to mixing. In each case, the excited mode is approximately 18-20dB pure relative to the background. See Supplementary §7 for details.

Conservation of light's OAM in air core fibers enables km-length-scale propagation of a large ensemble of spatial eigenstates, in analogy to the perturbation resistance of spinning tops and electron spin states. Therefore, this new photonic degree of freedom, having attracted much recent attention on account of its potentially infinite-dimensional basis, remains a conserved quantity over lengths practical for optical communications in appropriately designed fiber. Hence, we expect such fibers and their OAM states to play a crucial role in the general problem of increasing the information content per photon.

The authors would like to acknowledge J.Ø. Olsen for help with fiber fabrication, N. Bozinovic, S. Golowich and P. Steinvurzel for insightful discussions, and M. V. Pedersen for help with the numerical waveguide simulation tool. The authors gratefully acknowledge funding from the National Science Foundation (NSF) (ECCS-1310493, DGE-1247312) and DARPA (W911NF-12-1-0323, W911NF-13-1-0103).

See Supplement Materials below for supporting content

# Supplementary Materials

This section provides supplementary information to "Conservation of orbital angular momentum in air core optical fibers." The supplementary material is organized as follows. In §1 we list in detail the materials used for these experiments. In §2-3 we briefly review fiber vector and OAM modes, and fiber mode coupling. In §4 we provide additional detail on our fiber design methodology. In §5 we describe spiral interference images as a diagnostic tool. In §6 describe in detail our experimental OAM excitation scheme. In §7 we provide additional time domain data and describe measurement uncertainties.

## 1. Materials and Methods

The fiber index profile in **Fig. 1b** was obtained using an interferometry-based fiber profiler (IFA-100 Interfiber Analysis) at 633nm and is plotted with respect to the refractive index of silica. Since the profiler detects changes in the index of refraction interferometrically, an abrupt transition between air and glass is too large to be correctly calculated. Thus, the fiber is cleaved and allowed to soak in, via capillary action, index-matching oil (Cargille Laboratories) with $n \sim n_{Silica}$ to minimize the index discontinuity at that interface. To obtain a more accurate value of the radius of the air core for use in simulation, standard optical microscopy is applied along with a circle-fitting technique. SMF-28e is used as a scale reference. Simulations were performed with a numerical finite-difference method waveguide solver [1].

The fiber is cleaved with a Fujikura CT-100 fiber cleaver at tension ~200 grams. Care is taken in cleaving the fiber as cleaving at an angle will result in a discontinuity where the cleave's shearing action meets across the air core opposite to the cleaver blade. In phase, this discontinuity resembles an OAM perturbation and should be avoided.

The CW ECL used is a continuously tunable HP 8168F. The picosecond laser used is a PriTel FFL passively mode-locked fiber laser with a 3dB bandwidth of 0.6nm. Conventional SMF-28e is used for all single mode fiber links. The air core fiber is held on a Thorlabs Nanomax stage. Input coupling is performed with an f~4.5mm mid-IR coated aspheric lens.

The CPBS is comprised of a zero-order quarter wave plate and a polarization beam displacing prism. The camera is an InGaAs Allied Vision Technologies "Goldeye" camera. The fast detector is a New Focus 1444-50 picosecond photo-detector. The oscilloscope used was an Agilent infiniium DCA wide bandwidth oscilloscope, 86100A, with detector unit HP 83485B, 40 GHz port. The SLM used was a Hamamatsu LCOS-SLM x10468-08, with 600x800 pixels of pitch 20μm.

## 2. Vector and OAM Modes

Conventionally, for azimuthal mode number L > 1, fiber vector modes are designated as:

$$HE_{L+1,m}^{e,o} = F_{L,m}(r)\begin{Bmatrix}\hat{x}\cos(L\varphi) - \hat{y}\sin(L\varphi)\\ \hat{x}\sin(L\varphi) + \hat{y}\cos(L\varphi)\end{Bmatrix}e^{ik_{z,HE}\cdot z} \qquad \text{(S1)}$$

$$EH_{L-1,m}^{e,o} = F_{L,m}(r)\begin{Bmatrix}\hat{x}\cos(L\varphi) + \hat{y}\sin(L\varphi)\\ \hat{x}\sin(L\varphi) - \hat{y}\cos(L\varphi)\end{Bmatrix}e^{ik_{z,EH}\cdot z} \qquad \text{(S2)}$$

$F_{L,m}(r)$ is the mode's electric field envelope, where m is the radial mode number and m−1 the number of zeroes in F, $(x,y)$ and $(r,\varphi)$ are transverse coordinates, and $k_z$ is the mode's longitudinal wave vector, related to the effective index by $k_z = 2\pi n_{eff}/\lambda$. The HE[e,o] (even and odd) modes are degenerate with each other in circular fibers, as the vector fields are related by a rotation. Similarly, the EH[e,o] modes are degenerate; however, the pairs are not degenerate with each other, $k_{z,HE} \neq k_{z,EH}$.

The HE/EH polarization structure is spatially varying, as shown in **Fig. S1(a-b)**, and difficult to excite. An equivalent representation considers combinations of *degenerate* vector modes:

$$V_{L,m}^{\pm} = HE_{L+1,m}^{e} \pm iHE_{L+1,m}^{o} = \hat{\sigma}^{\pm}F_{L,m}(r)e^{\pm iL\varphi}e^{ik_{z,HE}\cdot z} \qquad \text{(S3)}$$

$$W_{L,m}^{\pm} = EH_{L-1,m}^{e} \mp iEH_{L-1,m}^{o} = \hat{\sigma}^{\pm}F_{L,m}(r)e^{\mp iL\varphi}e^{ik_{z,EH}\cdot z} \qquad \text{(S4)}$$

Here, $\hat{\sigma}^{\pm} = \hat{x} \pm i\hat{y}$ indicates left and right handed circular polarizations. We define an orbital angular momentum (OAM) state of topological charge $\mathcal{L}$ as:

$$\psi_{\mathcal{L}} = \hat{e}G(r)e^{i\mathcal{L}\varphi} \qquad \text{(S5)}$$

where $\hat{e}$ is an arbitrary but spatially uniform polarization, and $G(r)$ the field's radial envelope. It is evident from (Eqn. S 3-5) that these fiber modes are OAM states, with $\mathcal{L} = \pm L$ for $V_{L,m}^{\pm}$ and $\mathcal{L} = \mp L$ for $W_{L,m}^{\pm}$, and have spatially uniform polarizations (**Fig. S1 c**). The two kinds of OAM states, $V_{L,m}^{\pm}$ and $W_{L,m}^{\pm}$, are distinguished by comparing the handedness of OAM and circular polarization, with $V_{L,m}^{\pm}$ possessing spin and orbital angular momentum of the same handedness, and $W_{L,m}^{\pm}$ the opposite handedness. We refer to these two classifications as "spin-orbit aligned" and "spin-orbit anti-aligned," respectively. We define the 4 modes, $V_{L,m}^{\pm}$ and $W_{L,m}^{\pm}$, of the same (L,m) as an "OAM family." The HE/EH modes and the OAM modes, are *equivalent* descriptions of fiber eigenmodes of the vector wave equation under the weak-guidance approximation[2].

For fibers with small refractive index differences between core and cladding, $n_{eff_{HE}} \approx n_{eff_{EH}}$. In this case, an OAM family is susceptible to strong mode coupling from birefringent perturbations (detailed further in §3) and the fiber output typically resembles scalar linearly polarized (LP) "modes" which have a characteristic "beady" intensity pattern (see **Fig. S3 c**). However, the resulting field is not a fiber mode as it does not possess one distinct effective index. The splitting between degenerate pairs of OAM modes may be enhanced by designing a waveguide such that a large field coincides with a large index contrast [3]. Thus, ring fibers are advantageous, as they mimic the field structure of OAM modes. In addition the ring thickness can be chosen to restrict guidance to primarily the m=1 radial order, thus sculpting the density of states for guided modes.

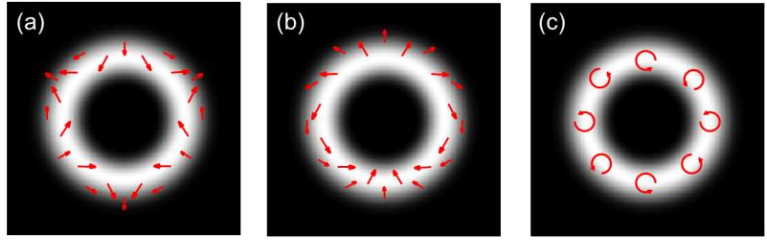

Fig. S1 Plot of the electric field of (a) the $HE^o_{31}$ and (b) $EH^o_{11}$ mode in an air core fiber (similar polarization patterns exist for any weakly-guiding circular fiber). (c) Plot of the electric field of an OAM mode made of complex combinations of the $HE_{31}$ even and odd modes. Note that the complicated spatially-dependent polarization has given way to a spatially uniform circular polarization, and this now resembles a free-space Laguerre-Gaussian beam.

## 3. Brief overview of fiber mode coupling

For any realistic waveguide, coupling between bound modes is expected, due either to deterministic perturbations or random fluctuations in the fiber. Following Bjarklev [4], we write the refractive index profile of a fiber as $n(r,z) = n_o(r) + n_b(r,\varphi,z)$, where $n_o$ is the fiber's ideal profile and $n_b$, which is separable into transverse and longitudinal parts, the latter denoted $f(z)$. The coefficient of power coupling between modes $j$ and $k$ is then determined by:

$$\langle P_{j,k}\rangle = \frac{\omega^2}{c^2}\Phi(k_{z,j}-k_{z,k})\left(\iint r dr d\varphi N_b E_j^* E_k\right)^2 \quad \text{(S6)}$$

where $\omega$ and c are the frequency and speed of light, $\Phi$ is the spatial power spectrum of the autocorrelation of f (z), $E_j$ is the normalized electric field of the jth mode, and $N_b(r, \varphi) = n_b(r,\varphi,z)/f(z)$ is the transverse perturbation. If the autocorrelation of f(z) is assumed to be of Gaussian form, then:

$$\Phi(k_{z,j}-k_{z,k}) = \sqrt{\pi}\sigma^2 L_c e^{-\left[\frac{1}{2}L_c(k_{z,j}-k_{z,k})\right]^2} \quad \text{(S7)}$$

where $\sigma$ is the RMS deviation of $f(z)$, and $L_c$ is its correlation length.

The angular part of the perturbation can be separated by

$$N_b(r,\varphi) = \sum_{p=-\infty}^{\infty} \widetilde{N_p}(r) e^{ip\varphi} \quad \text{(S8)}$$

If the j$^{th}$ and k$^{th}$ modes are OAM states, as specified by (S 5), we can separate the integrals in $(r,\varphi)$, and simplify the angular part of the integral in (S 6):

$$\iint r dr d\varphi N_b E_j^* E_k \propto \sum_{p=-\infty}^{\infty} \int_0^{2\pi} d\varphi e^{i(\mathcal{L}_k-\mathcal{L}_j+p)\varphi} \quad \text{(S9)}$$

(S 9) vanishes unless

$$p - (\mathcal{L}_j - \mathcal{L}_k) = 0 \quad \text{(S10)}$$

This selection rule implies overall conservation of orbital angular momentum. Knowledge of the angular Fourier series coefficients [5], $\widetilde{N_p}$, enables prediction of the results of a perturbation. For instance, it is evident that for an untitlted fiber bragg grating no transfer of OAM between the input mode and perturbation should be possible, since $\widetilde{N_p} = 0 \ \forall p \neq 0$, although if the grating is tilted, transfer is permitted.

In (S 7), since $\Phi$ achieves maximum values for $\Delta n_{eff} = n_{eff\,j} - n_{eff\,k} = 0$, maximum exchange of power between modes will be achieved when they are phase matched. In theory, two degenerate modes will never couple if the integral in (S6) is zero. In practice any real perturbation will have some Fourier spectrum in both azimuthal and radial coordinates, as in equation (2); thus, coupling between degenerate states is generally expected. A simple manifestation of this is coupling between the two polarization modes in conventional single mode fibers because of which polarization cannot be maintained. The question becomes one of coupling length scale. In view of phase matching considerations only, strong coupling between the spin-orbit aligned states, $V_{L,m}^{+}$ and $V_{L,m}^{-}$, would be expected. However, to achieve this coupling a perturbation, $n_b$, must couple *both* opposing SAMs and OAMs of arbitrary $|L|$, in order to conserve total angular momentum flux. As indicated in **Fig 3a**, such a perturbation is unlikely for high $|L|$. In view of coupling strength relative to the integral in (S6), coupling is likely between the modes $V_{L,m}^{+}$ and $W_{L,m}^{-}$, since any birefringent perturbation will be sufficient to couple $\hat{\sigma}^{+}$ and $\hat{\sigma}^{-}$. Indeed, this coupling is evident in even a few meters of conventional fibers. However, for OAM states in our air core fiber, this coupling is inhibited because states of the same $L$ are non-degenerate.

## 4. Modes of air core fibers

Desired OAM states possess quantum numbers (L,m) = ($L$,1). Modes with higher radial quantum numbers should be suppressed because they can lead to so-called "accidental degeneracies," in which a mode with quantum numbers L,M may be degenerate, over some wavelength range, with a mode with quantum numbers L', M', where L and L' may be several integers apart (**Fig. S2 a**; note that modes of different azimuthal symmetry have no avoided-crossings that would prevent these accidental degeneracies). Coupling between such states necessitates a perturbation with OAM content $L$'- $L$ which, although still potentially large, is generally more feasible than a perturbation with content $2L$, which will destroy a fiber's ability to transmit higher-order OAM states over long distances.

The most obvious design methodology for suppressing m>1 coupling is to make a thin, high-contrast ring, for which the m>1 states simply are not guided. Recent work has theoretically proposed such waveguides [6], and experimental work has claimed OAM propagation over cm-lengths [7]. However, when the index-ring is made very thin, differences in electromagnetic continuity conditions between **s** and **p** polarized electromagnetic fields will cause spin-orbit coupling, a fundamental effect preventing km-scale OAM state propagation in such thin-ring air-core fibers [8]. In order to avoid this debilitating effect, we allow for a thicker ring, for which the presence of m>1 modes may be unavoidable. In such waveguides, the additional design requirement is the avoidance of accidental degeneracy in the wavelength region of interest. Examples of m=1 and m=2 modes in the air core fiber are shown in **Fig. S2 b** and **2 c**, where we wish to avoid operation in regions in which the m=2 modes are closely spaced in $n_{eff}$ to desired OAM modes.

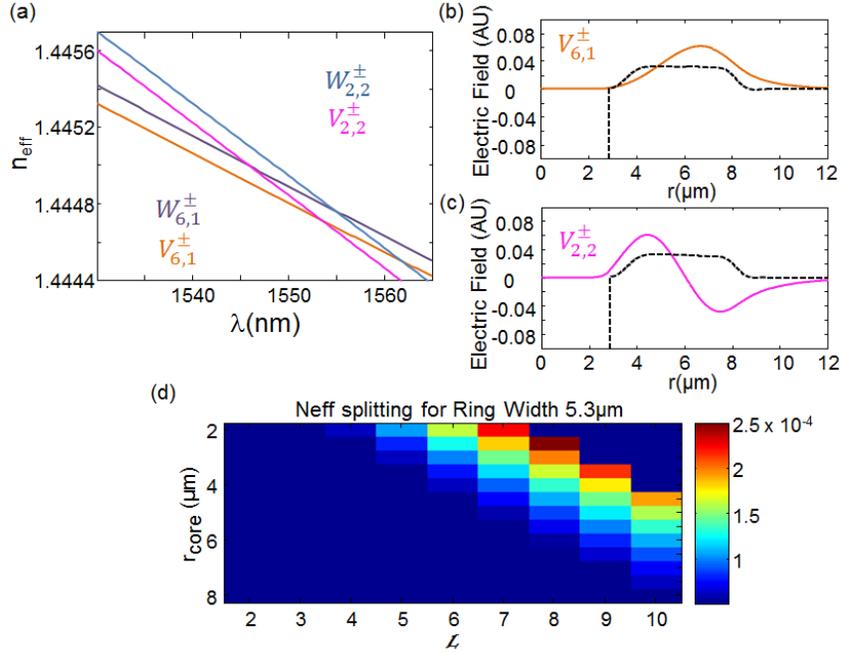

Fig. S2 (a) Accidental degeneracy in a particular fiber design between the (L,m) = (6,1) and (2,2) modes. Even though the $|\mathcal{L}|=6$ OAM modes are sufficiently split in effective index, they are degenerate with another OAM family, preventing pure propagation of either family. (b) Radial envelope for the transverse field, $E^t(r)$, of one of the $|\mathcal{L}|=6$ OAM modes. The radial envelopes for every mode within an OAM family are nearly identical. (c) Radial envelope of the mode field of an m=2 mode. Due to the possibility for accidental degeneracy these modes should be avoided, although they are valid OAM states. (d) $n_{eff}$ splitting as a function of $|\mathcal{L}|$ and $r_{core}$ for a constant ring width. Depending on the chosen radii, an ensemble of OAM states with different members is supported. This forms a band OAM states as a function of $r_{core}$, and is generally designable as desired.

The design yields a band of OAM states in $|\mathcal{L}|$, for which the $n_{eff}$ splitting values are large enough ($\geq$1e-4) to enable propagation of pure OAM states (see discussion in [3] for a phenomenological reasoning behind this difference as a near-universal metric in optical fibers). Below the lowest $|\mathcal{L}|$ for which OAM states are supported, the $n_{eff}$ splitting is too small and LP-like behavior is obtained. Above the largest $|\mathcal{L}|$, modes are cut-off. The specific set of $|\mathcal{L}|$ states which support OAM propagation are a function of the dimensions of the air core fiber: the outer radius of the air core, $r_{core}$, outer radius of the ring, $r_{ring}$, and the index contrast between the ring and the silica cladding, $\Delta n$. This waveguide design optimization is illustrated in **Fig. S2 d** for fiber with large index contrast, $\Delta n = 0.035$. As $r_{core}$ is increased for constant ring widths ($r_{ring} - r_{core}$), the mode volume increases and higher $|\mathcal{L}|$ OAM states are supported. A similar effect is observed when mode volume is changed by allowing $r_{ring}$ to vary. However, as mode volume is increased, the number of m=2 modes also increases in a nonlinear manner, thus, $r_{core}$ and $r_{ring}$ must be chosen in tandem to support the desired OAM ensemble while avoiding accidental degeneracy. For the fabricated air core fiber tested in this work, $r_{core}$=3μm and $r_{ring}$=8.25 μm, designed for an OAM ensemble of $|\mathcal{L}| = \{5,6,7\}$ in the 1550nm wavelength range.

## 5. Brief overview of OAM-Gaussian (spiral) interference

**Fig. S3 a** depicts a simulated interference between a pure, defocused $L = -6$ OAM state and a Gaussian beam. The right-handed spiral indicates, in the convention we have chosen, $L < 0$, while the six spiral arms, or parastiches, indicates the absolute value of $L$. Comparing the simulated behavior shown in **Fig. S3 a** with that shown in **Fig. S3 b**, we observe that when the OAM state degrades in purity by the addition of a meager amount (~ −14dB) of $L = 6$, the arms become "beady." In the case where $L = 6$ and $L = -6$ have equal contributions (**Fig. S3 c**), the arms give way entirely to a circularly symmetric constellation of $2|L|$ beads, somewhat similar to the field of an LP state; this is indeed the origin of observation of only LP-like states in conventional fibers, where mixing between all modes of the same OAM family readily occurs. If an OAM state is degraded by the presence of a neighboring $L$ state (**Fig. S3 d**), the intensity distribution of the interference pattern becomes asymmetric, although the number of spiral arms is unchanged. One expects the same effect to occur in an interference between a pure OAM mode and an off-centered Gaussian. In contrast to the degraded behaviors discussed above and shown in **Fig. S3**, we find that the experimentally recorded images in **Fig. 2** demonstrate clean OAM state propagation in air core fibers.

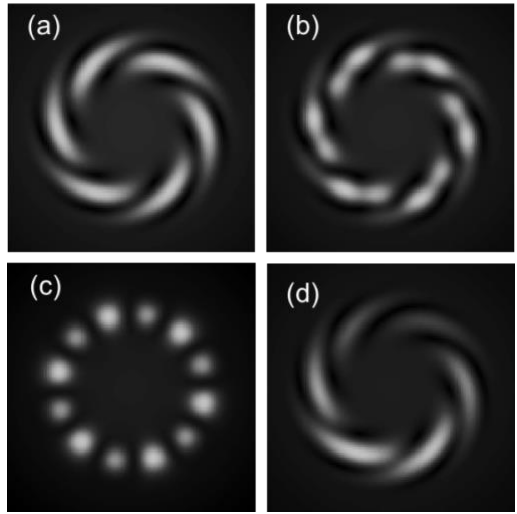

Fig. S3 (a) Simulated interference pattern between an expanded Gaussian and an $L = -6$ OAM mode. The number of spiral arms indicates $|L|$ for a pure OAM state, and the handedness indicates whether $L$ is positive or negative. (b) Interference pattern between a Gaussian and an OAM beam with dominant component $L = -6$ and ~14dB MPI into $L = 6$. Such impurity is manifest in beadiness along the spiral arms. (c) Interference between a Gaussian beam and a beam with equal parts $L = 6$ and $L = -6$. Spiral arms give way to a LP mode-like constellation. (d) Gaussian interference with a beam containing $L = -6$ with 6dB MPI in $L = -5$. Note that the same effect is expected if the Gaussian is off-center.

## 6. Experimental coupling scheme

Excitation of fiber OAM modes requires matching one or a set of free-space OAM states to the corresponding fiber state in size and relative dimensions for a variety of $L$, and ensuring that the incident beam is both tilt-free and on-axis relative to the fiber to avoid projection into other fiber OAM states. To achieve our goal we use two holograms. A collimated Gaussian beam illuminates the first pattern, SLM1 (equipment details in §1), on which is displayed:

$$SLM1(x,y,r,\varphi) = \left(\mathcal{L}\varphi + k_{SLM}x + \frac{r^2\pi}{\lambda F} + M(r,\varphi)\right) * LG_{\mathcal{L},0}(r;w) \tag{S11}$$

Here $(x,y) = (m,n)*\Delta x$ are the SLM's spatial coordinates, which are equal to some integer (m,n) multiples of the SLM pixel pitch, $\Delta x$, where $m=n=0$ indicates the SLM's center. $(r,\varphi)$ are the (approximate) polar coordinates of the SLM, $k_{SLM}$ is a blazed grating wave vector of programmable magnitude, $\lambda$ is the wavelength of operation, F is the focal length of some lens to which the phase curvature (third term in brackets in (S11)) corresponds, $LG_{\mathcal{L},0}(r;w)$ is the radial envelope of the $LG_{\mathcal{L},0}$ Laguerre Gauss mode of beam waist, $w$, scaled such that the peak amplitude is 1, and $M(r,\varphi)$ is a phase function related to the $LG_{L,0}$ mask. The azimuthal phase, in conjunction with the phase mask, M, and amplitude mask, $LG_{\mathcal{L},0}$, optimize coupling into the $LG_{\mathcal{L},0}$ mode in the far field [9,10]. Since this conversion happens with sub-optimal efficiency, a phase tilt, $k_{SLM}x$, is added to spatially isolate the desired diffraction order. Additionally, a phase curvature corresponding to F is added, such that a distance F from the SLM, the far-field is obtained as though the beam were passed through a lens. In our experiments, F~1m.

The diffracted beam, a distance F from SLM1, is multiplied by a second phase hologram:

$$SLM2(x,y,r,\varphi) = \left(\frac{r^2\pi}{\lambda F_2}\right)circ\left(\frac{r}{R}\right) + attn * \left(1 - circ\left(\frac{r}{R}\right)\right) \tag{S12}$$

Where $F_2$ is, again, the focal length of the lens with phase curvature corresponding to the first term in (S12), and R is a scale factor determined *in situ*. This pattern is designed to, within the region of interest $r < R$, multiply by a spherical phase. Outside this region of interest, an attenuation pattern, *attn*, is applied, such that light in the *attn* region is scattered in all directions [11]. This completely separates the first diffracted order of SLM1 from the rest. In our experiments, R~2mm, dependent on F and $w$ in (S 11).

The near field of SLM2 is then imaged by an f~4.5mm aspheric lens into the fiber under test, through a zero-order quarter wave plate (aligned to negligibly degrade or displace the incident OAM state). $F_2$ is chosen to offset the additional phase curvature accrued by such an imaging operation. Higher diffracted orders from SLM2 are vignetted, either by the clear aperture of the coupling lens, or by the air core fiber itself.

The primary advantage of this method is as follows: in free space, the ring radius of an LG beam after focus is determined by the beam's topological charge, $\mathcal{L}$, initial beam waist, and lens focal length. Thus, beams of different $\mathcal{L}$ map to different sizes in the focal plane of a given imaging system. In the air core fiber, however, modes with different $\mathcal{L}$ are constrained by the waveguide to have roughly the same radius. Thus, by performing only a focusing operation, beam size and lens focal length cannot be simultaneously optimized for coupling different $|\mathcal{L}|$ into the air-core fiber. The two-stage excitation system, on the other hand, enables exciting a wider range of desired OAM states of the fiber by digitally manipulating the SLM parameters.

We have found that the shape of the radial envelope is less critical to mode excitation compared with beam axis-overlap. If the fiber is offset from the incident beam by distances on the order of 0.1μm, mode purity can drop by an order of magnitude (10dB). Similarly, tuning the incident wavelength without changing the phase tilt, $k_{SLM}$, on SLM1 by as little as 5nm can result in input coupling mode purity degradation of ~10dB.

While our coupling scheme is necessarily component intensive and alignment sensitive, it suffices for the purposes of the primary research goals herein – of studying fiber propagation and stability of OAM. Alternative coupling/multiplexing strategies include OAM mode sorters [12], integrated silicon photonic interferometric couplers [13], and cascaded dove prism interferometers [14], and we note that the input/output coupling problem is beginning to receive greater attention from the community, as practical applications of OAM, including using fibers that can support them, develop.

## 7. Supplementary time domain measurements and error estimation

For the time domain measurements described in discussion related to **Fig. 4**, data was taken for all 8 modes of the $|L| = 5$ and $|L| = 6$ OAM families. Data for the $L = -5$ and $L = -6$ states is shown in **Fig. S4 a**, with individual traces in **Fig. S4 b-d**. Each mode overlaps in time with its degenerate partner from **Fig. 4**, and exhibits similar MPI compared to states shown in the main text as expected; the sign of $L$ should not alter mode purity.

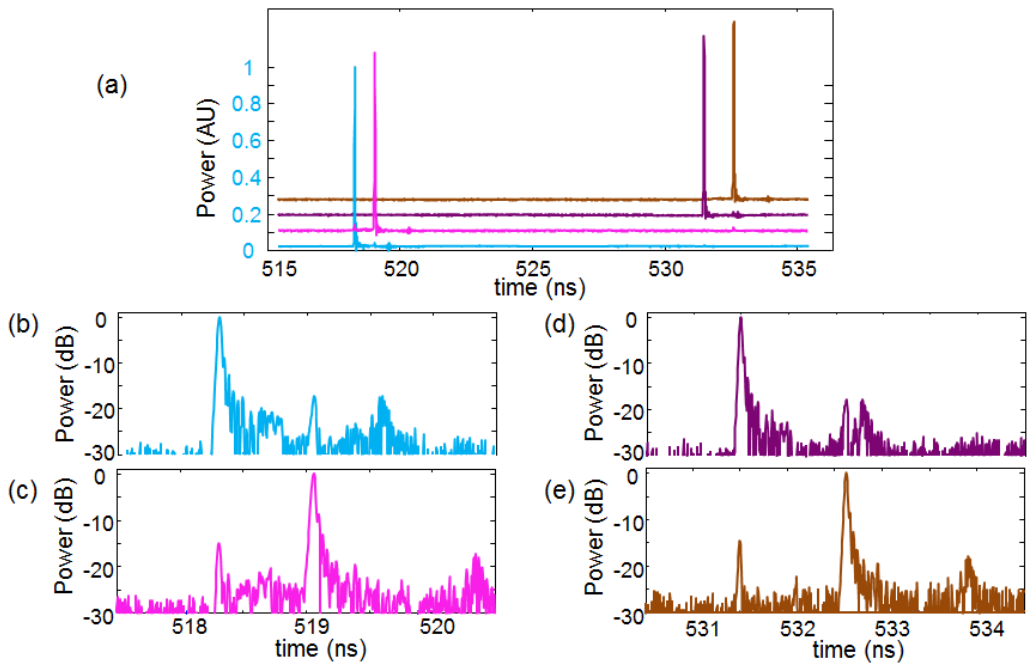

Fig. S2 (a) Time domain traces for the remaining 4 OAM modes in the $|L| = 5$ and $|L| = 6$ OAM families not shown in Fig 5a. Traces are offset for vertical clarity, in order of increasing group delay. Each mode overlaps in time with its degenerate partner, within the resolution limit of the measurement, and has comparable MPI. (b-e) Close-ups of time domain trace for $L = -5\ \hat{\sigma}^+$, $L = -5\ \hat{\sigma}^-$, $L = -6\ \hat{\sigma}^+$, and $L = -6\ \hat{\sigma}^-$. In each case, the excited mode is at least 15dB pure in relation to the background.

The reason that simple peak power ratios obtained from the temporal traces suffice in calculating mode purity or MPI is that the high $|\mathcal{L}|$ OAM states in air core fibers have similar group-velocity dispersion values. The dispersion values, as simulated from the measured refractive index profile of the air-core fiber, are 60.0 and 58.0 ps/nm/km for the spin-orbit aligned and anti-aligned modes of the $|\mathcal{L}| = 5$ family, and 64.3 and 62.4 ps/nm/km for the aligned and anti-aligned modes of the $|\mathcal{L}| = 6$ OAM family. So, our ~7ps pulsed source (FWHM ~ 0.6nm), broadens to similar output pulse widths of 36.7 and 35.5ps for the |L|=5 spin-orbit aligned and anti-aligned modes, and 39.2 and 38.0ps for the |L|=6 aligned and anti-aligned modes, respectively, resulting in similar peak power reductions after 1km of propagation.

One limit to this time-of-flight measurement is laser pulse drift in time with respect to the electrical trigger signal on the order of 10s of picoseconds (**Fig. S5 a**). Whether this drift is intrinsic to the laser or is a consequence of thermal/environmental fluctuations of the optical system as a whole is unclear. The spurious peak shown approximately 1.5ns after each main measurement peak is a result of electrical ringing in the detector + oscilloscope receiver. **Fig. S5 b** shows similar impulse response behavior for pulsed transmission through 2m of single mode fiber at 1550nm, in which this behavior cannot be a result of multimode effects in the fiber under test. Thus, we conclude that the spurious peaks must be considered part of the measurement background.

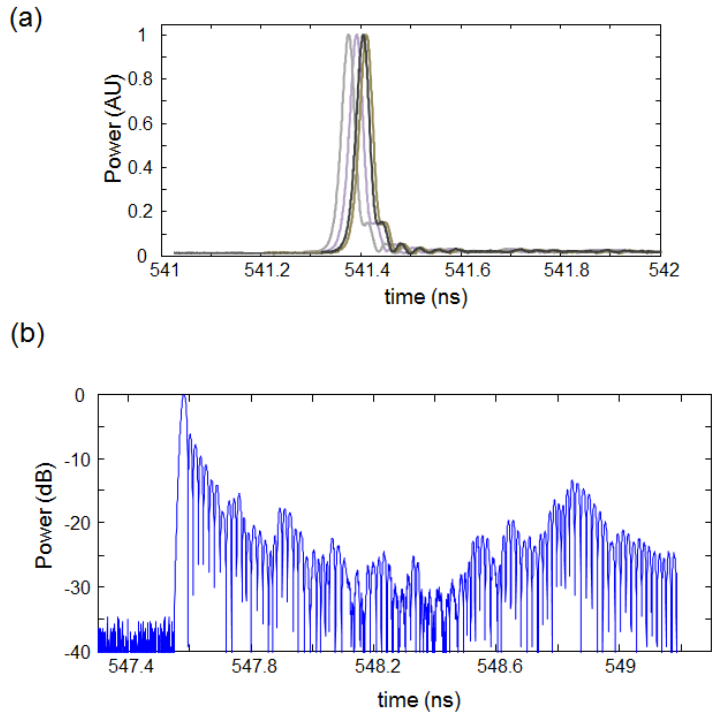

Fig. S5 (a) Several traces of the same state over a period of a few minutes. Drift from peak to peak provides an accuracy limitation for this measurement and may be caused by fluctuations inside the laser relative to its electrical trigger, or from temperature/environmental fluctuations of the optical system as a whole. (b) Time of flight measurement for transmission over 2m of single mode fiber, plotted on log scale. The ringing around 548.9ns is part of the impulse response of the receiver, and is evident (around 1.5ns after the main peak) in all time of flight measurements.

Additionally, the detector for these measurements is polarization-sensitive on the level of about 8~10%. Assuming the worst case error (the parasitic mode is detected 10% less efficiently than the signal mode, or vice versa), this leads to an error in measured mode purity of ±0.5dB.